\documentclass[aps,prl,reprint,superscriptaddress]{revtex4-2}
\usepackage{graphicx}   
\usepackage{mathtools}
\usepackage{mathrsfs}
\usepackage{amsmath}
\usepackage{dcolumn}    
\usepackage{bm}         
\usepackage{hyperref}   
\usepackage[mathlines]{lineno}
\usepackage{float}
\usepackage{xfrac}
\begin{document}
\title{Silicon charge pump operation limit above and below liquid helium temperature}
\author{Ajit Dash}
\email{ajit.dash@unsw.edu.au}
 \affiliation {School of Electrical Engineering and Telecommunications, University of New South Wales, Sydney, NSW 2052, Australia} 
\author{Steve Yianni}
\affiliation{School of Electrical Engineering and Telecommunications, University of New South Wales, Sydney, NSW 2052, Australia}
 \affiliation {Diraq, Sydney, NSW 2052, Australia}
 \author{MengKe Feng}
\affiliation{School of Electrical Engineering and Telecommunications, University of New South Wales, Sydney, NSW 2052, Australia}
 \affiliation {Diraq, Sydney, NSW 2052, Australia}
\author{Fay Hudson}
\affiliation{School of Electrical Engineering and Telecommunications, University of New South Wales, Sydney, NSW 2052, Australia}
 \affiliation {Diraq, Sydney, NSW 2052, Australia}
 \author{Andre Saraiva}
\affiliation{School of Electrical Engineering and Telecommunications, University of New South Wales, Sydney, NSW 2052, Australia}
 \affiliation {Diraq, Sydney, NSW 2052, Australia}
\author{Andrew S. Dzurak}
\affiliation{School of Electrical Engineering and Telecommunications, University of New South Wales, Sydney, NSW 2052, Australia}
 \affiliation {Diraq, Sydney, NSW 2052, Australia}
\author{Tuomo Tanttu}
 \email{t.tanttu@unsw.edu.au}
\affiliation{School of Electrical Engineering and Telecommunications, University of New South Wales, Sydney, NSW 2052, Australia}
 \affiliation {Diraq, Sydney, NSW 2052, Australia}
\date{\today}
\begin{abstract}

Semiconductor tunable barrier single-electron pumps can produce output current of hundreds of picoamperes at sub~ppm precision, approaching the metrological requirement for the direct implementation of the current standard. Here, we operate a silicon metal-oxide-semiconductor electron pump up to a temperature of 14~K to understand the temperature effect on charge pumping accuracy. The uncertainty of the charge pump is tunnel limited below liquid helium temperature, implying lowering the temperature further does not greatly suppress errors. Hence, highly accurate charge pumps could be confidently achieved in a $^4$He cryogenic system, further promoting utilization of the revised quantum current standard across the national measurement institutes and industries worldwide.
\end{abstract}
\maketitle

The seven base International Systems of Units (SI) serve as basis for measuring any physical quantity. Redefining these units over the years aims to ascertain a consistent and universal metrological standard. Recent revision of SI suggests the use of quantized charge pump for a practical realization of the primary current standard, by agreeing a fixed value of elementary charge $e$ $(=1.602176634 \times 10^{-19}$~A$\cdot$~s)~\cite{mohr2018data, rossi2021single}. A charge pump is a nanoelectronic device that transfers integer $n$ number of electrons, holes or cooper pairs per voltage cycle with frequency $f$, yielding quantized current $I$ $(=\textit{n}\times\textit{e}\times\textit{f})$. A clock-controlled on-demand charge emitting characteristics of charge pump also attracts attention in the field of quantum information processing and quantum optics~\cite{johnson2018phonon, yamahata2019picosecond}. Significant research has been pursued by the national measurement institutes and academia to realize quantized pumping of quasi-particles in variety of metal, superconductor, metal-superconductor hybrids and semiconductor systems~\cite{scherer2019single}. 

Silicon metal-oxide-semiconductor (SiMOS) nanostructure based charge pumps has evinced the potential of practical realization of the SI Ampere by demonstrating remarkable combination of pumping speed and fidelity~\cite{zhao2017thermal, giblin2020realisation, yamahata2017high, fujiwara2008nanoampere, yamahata2015gigahertz, rossi2018nanolett}. Besides, quantum devices fabricated on Si have exhibited significant reduction of $1/f$ noise and background charge fluctuations at regimes of high amplitude operations~\cite{koppinen2012fabrication, zimmerman2008long}. SiMOS gate-stack technology enables fabrication of multi-layer top gates, facilitating strong planar electrostatic confinement to define a quantum dot (QD)~\cite{rossi2014accurate}. Owing to the small physical size, the electronically defined QD in Si has high charging energy, hence capable of transferring discrete number of charges at a base temperature of sample-space $(T_{\mathrm{base}})$ up to few kelvins~\cite{fujiwara2008nanoampere, yamahata2015gigahertz, rossi2018nanolett, giblin2020realisation, yamahata2016gigahertz}. However, the influence of temperature on the charge pumping accuracy has not been assessed in any physical system. Understanding the temperature limit is crucial to choose an optimal $T_{\mathrm{base}}$, which might further relax the requirement of $^{3}$He or dilution refrigerator. In this work, we realize quantized electron pumping in a voltage-induced Si QD up to a $T_{\mathrm{base}}$ of 14~K. Later, we fit the measured current plateaus to decay-cascade and thermal model of charge transfer to quantify the charge capturing mechanism as a function of $T_{\mathrm{base}}$ and periodic drive amplitude. We realize the transition temperature between tunnel limited and thermally limited electron capture mechanism in our Si based charge pump is higher than the liquid helium temperature ($\approx$~4.2~K) for a wide range of pulsing drive amplitude while pumping single-electron per voltage cycle of 100~MHz frequency. These results forecast the theoretical errors of the charge pump when operated above and below the transition temperature.

\begin{figure*}[ht!]
\includegraphics[width=\textwidth]{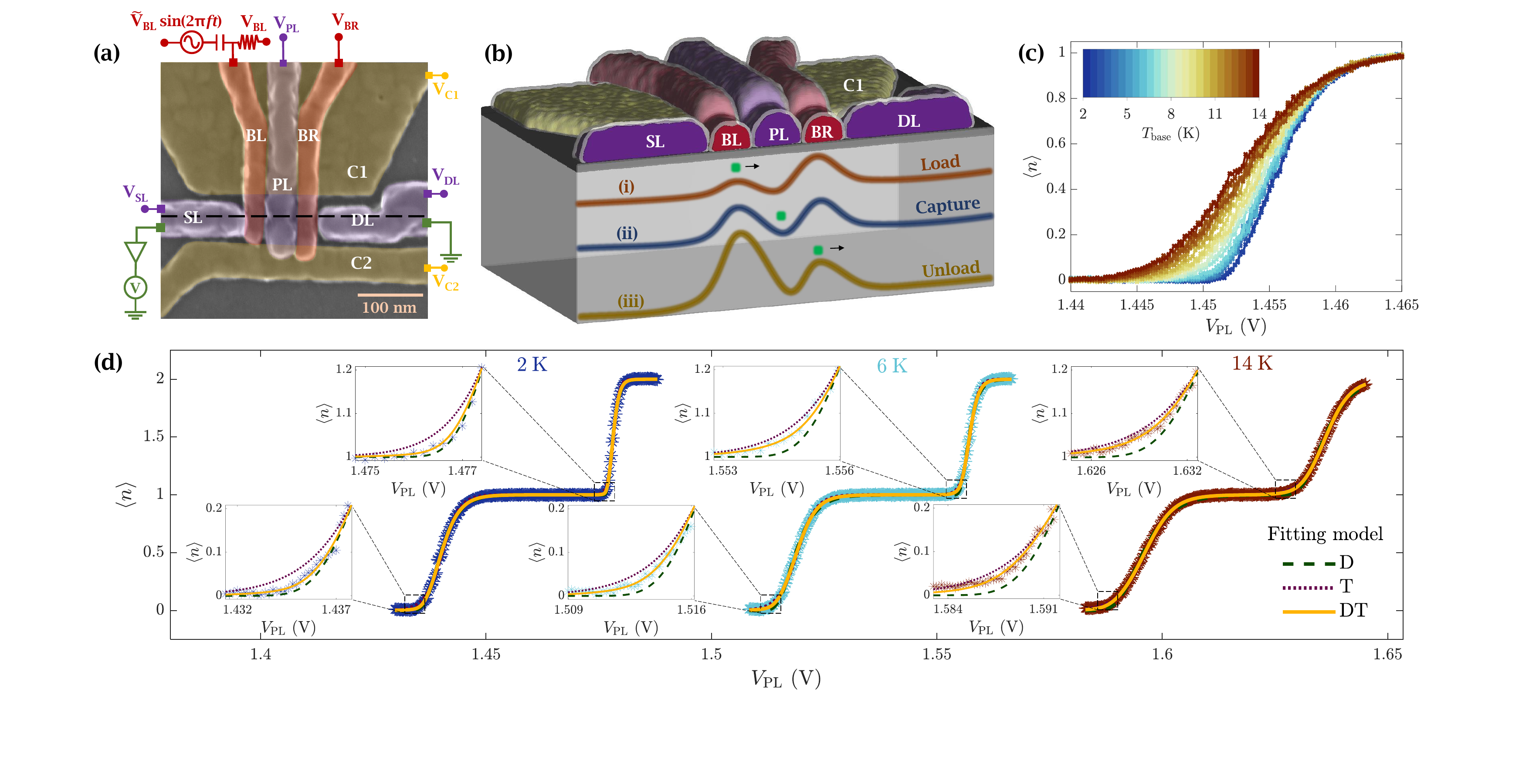}
\caption{\label{fig:schematic} (a) False-color scanning-electron-micrograph of an  electron pump similar to one used in the experiment together with schematic of the measurement setup. (b) Three-dimensional cross-section schematic of the charge pump  along the black dashed line showing the three layers (yellow: 20~nm, red: 27~nm and violet: 35~nm) gate-stack architecture. (i) load, (ii) capture and (iii) unload illustrate conduction-band energy level profile during three stages of an electron (green dot) pumping cycle. (c) Average number of pumped electrons per AC voltage cycle $\langle n \rangle$ as a function of plunger gate voltage $V_{\mathrm{PL}}$ with varying base temperature of sample-space $T_{\mathrm{base}}$ up to 14~K. (d) Plateaus of measured $\langle n \rangle$ at $T_{\mathrm{base}}$ 2~K, 6~K, and 14~K  along with its fit to decay-cascade model (D), thermal model (T), and weighed sum of decay-cascade and thermal model (DT) of charge pumping. Insets show the zoomed-in axes at the raising edge of first and second plateaus. The data is horizontally shifted for clarity.}
\vspace{-1.5em}
\end{figure*}

The charge pump measured in this work is fabricated on a near-intrinsic silicon substrate with a 7~nm thick thermally grown silicon dioxide (SiO$_2$) layer. The aluminium (Al) gate-stack architecture in Figs.~\ref{fig:schematic}(a, b) is realized by defining the device morphology with electron-beam-lithography, followed by thermal evaporation of subsequent three Al metal layers. Al top-gates are electrically insulated from the adjoining metal gates by thermally growing aluminium oxide (Al$_{\mathrm{x}}$O$_{\mathrm{y}}$) of 3~nm thickness between each layer. Top-gates are connected to a programmable room-temperature DC bias source through 300~MHz cryogenic low-pass-filter. A QD is electrically induced under plunger gate (PL) by tuning the planar confinement potential $(V_{\mathrm{C1}}$ and $V_{\mathrm{C2}})$ and tunnel barrier potential $(V_{\mathrm{BL}}$ and $V_{\mathrm{BR}})$. Clock-controlled charge transfer characteristics of the pump is instigated by adding an AC $sine$ waveform with a time-period of 10~ns, generated using an arbitrary-waveform-generator connected to BL pulsing barrier gate in Fig.~\ref{fig:schematic}(a). The periodic drive modulates BL barrier potential as show in Fig.~\ref{fig:schematic}(b) to load electron from the source reservoir (i), followed by capturing the electron in the QD (ii) and finally unloading it to drain reservoir (iii), generating a pump current $(I)$. The output current is pre-amplified with a gain of $10^{8}$~V$/$~A using a transimpedence-amplifier and measured using a voltmeter by integrating over a time of 20~ms, or one power line cycle. The pumped current is normalized as $I/ef$ to elucidate the average number of electrons pumped per AC voltage cycle $\langle n \rangle$. All the measurements are performed in a variable-temperature-insert at $T_{\mathrm{base}}$ that ranges from 2~K to 14~K with a cryogenic temperature controller. A single sweep of measured $T_{\mathrm{base}}$ dependent normalized pump current as a function of plunger gate voltage $V_{\mathrm{PL}}$ at a constant AC periodic drive amplitude $\widetilde{V}_{\mathrm{BL}}$ of 350~mV is displayed in Fig.~\ref{fig:schematic}(c). 

\begin{figure*}[ht!]
\includegraphics[width=\textwidth]{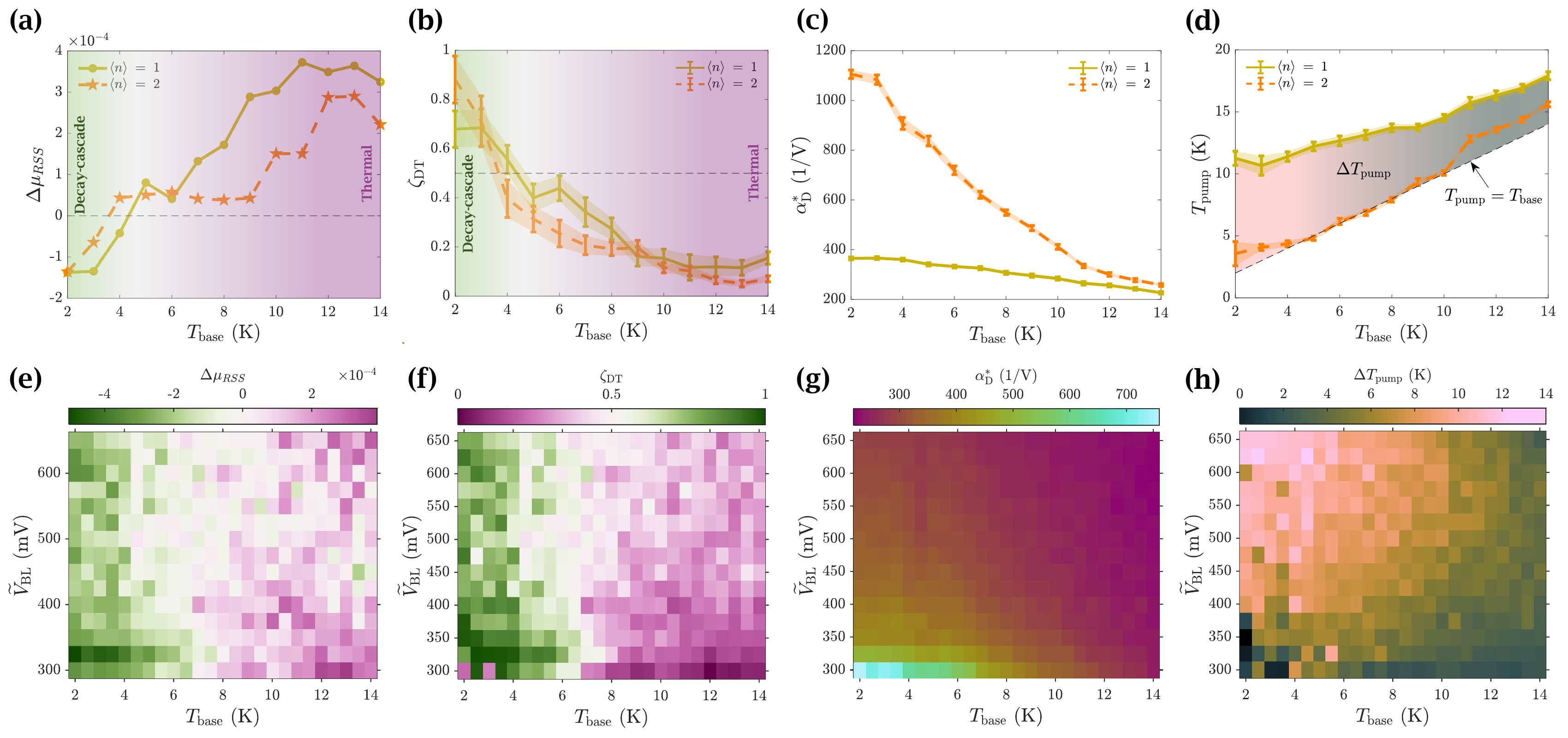}
\caption{\label{fig:1e&2e} Phenomenological fitting parameters along with error bars, extracted from the decay-cascade model (D), thermal model (T), and weighed sum of decay-cascade and thermal model (DT) charge pumping for first and second plateau as a function of base temperature of sample-space $T_{\mathrm{base}}$. (a) Difference between the average residual sum of squares of decay-cascade and thermal fits of the measured data $(\Delta \mu_{RSS})$. (b) Weight component of decay-cascade $(\zeta_{\mathrm{DT}})$, and thermal $(1- \zeta_{\mathrm{DT}})$ in the combined model. (c) Gate-referred tunnel rate factor $(\alpha^{*}_{\mathrm{D}})$. (d) Local electron temperature at the source and drain reservoir leads  $(T_{\mathrm{pump}})$. The difference between the black-dashed line $($at $T_{\mathrm{base}})$ and $T_{\mathrm{pump}}$ depict the AC periodic drive induced heating in the pump $(\Delta T_{\mathrm{pump}})$. (e) $\Delta \mu_{RSS}$, (f) $\zeta_{\mathrm{DT}}$, (g) $\alpha^{*}_{\mathrm{D}}$ and (h) $\Delta T_{\mathrm{pump}}$ for single-electron pumping, measured as a function of $T_{\mathrm{base}}$ and drive amplitude $(\widetilde{V}_{\mathrm{BL}})$.}
\vspace{-1.5em}
\end{figure*}

We observe smoothing rise to the $\langle n \rangle$ plateau with increasing  $T_{\mathrm{base}}$ from 2~K to 14~K, in Fig.~\ref{fig:schematic}(c). This corroborates the occurrence of dissimilar electron transfer mechanism in our charge pump, which is explained by decay-cascade~\cite{kashcheyevs2010universal} and thermal~\cite{yamahata2011accuracy, zhao2017thermal} models by elaborating the process of periodic decoupling of QD from source reservoir lead. The decay-cascade model assumes that the dominant charge pumping error mechanism is a series of non-equilibrium electron escape events back to the source reservoir to yield $\langle n_{\mathrm{D}}\rangle$ number of trapped electrons in the QD, given as: 
\vspace{-0.5em}
\begin{equation}
\langle n_{\mathrm{D}}\rangle = 
\sum_{i=1}^{2} \exp{\{-\exp{[\alpha^{*}_{\mathrm{D}_{i}}(V_{\mathrm{PL}}-V_{0_{i},\mathrm{D}})]}\}}
\label{eq:decaycascade}
\end{equation}
where $\alpha^{*}_{\mathrm{D}}$ is the gate-referred tunnel rate factor and $V_{0,\mathrm{D}}$ is the threshold voltage obtained from decay-cascade fit of the normalized pump current. The double-exponential function describing the decay-cascade regime in Eq. \ref{eq:decaycascade} analytically has an asymmetric rise shape.

At elevated temperatures, the broader energy spectrum of the electron reservoir becomes the dominating error process, to capture $\langle n_{\mathrm{T}}\rangle$ number of electrons in the QD~\cite{zhao2017thermal}. The average number of electrons pumped in the thermal regime is ascertained by the Fermi distribution of electrons in the source reservoir leads at thermal equilibrium, expressed as: 
\begin{equation}
\langle n_{\mathrm{T}}\rangle = 
\sum_{i=1}^{2} 1 / \{1+\exp{[\beta^{*}_{\mathrm{T}_{i}}(V_{\mathrm{PL}}-V_{0_{i},\mathrm{T}})]}\}
\label{eq:thermal}
\end{equation}
where $\beta^{*}_{\mathrm{T}}$ is the gate-referred modified thermodynamic beta and $V_{0,\mathrm{T}}$ is the threshold voltage obtained from the thermal fit of the normalized pump current, which analytically ascribe to a symmetrical rise shape. The phenomenological fit parameter $\beta^{*}_{\mathrm{T}}$ corresponds to heat induced by the AC periodic drive, required to prompt the charge pumping process, and inferred as $\beta^{*}_{\mathrm{T}} = (e \cdot \alpha_{\mathrm{PL-QD}})/(k_{B} \cdot T_{\mathrm{pump}})$. Here, $\alpha_{\mathrm{PL-QD}}=e(\Delta V_{\mathrm{SD}}/\Delta V_{\mathrm{PL}})$ is lever-arm of PL to pump QD and $T_{\mathrm{pump}}$ is the local electron temperature at the source and drain reservoir leads. We calculated $\alpha_{\mathrm{PL-QD}}$ from the slope of experimentally measured $V_{\mathrm{PL}}$ versus $V_{\mathrm{SD}}$ when an electron is added to the QD from the source reservoir.

We suspect involvement of both the non-equilibrium (decay-cascade) and equilibrium (thermal) charge capturing mechanism in the pumping process. Therefore, we propose a weighed sum of decay-cascade and thermal model, quoted as combined model, given as:
\begin{multline}
\langle n_{\mathrm{DT}}\rangle = 
\sum_{i=1}^{2} (\zeta_{\mathrm{DT}_{i}}) \cdot \exp{\{-\exp{[\alpha^{*}_{\mathrm{DT}_{i}}(V_{\mathrm{PL}}-V_{0_{i},\mathrm{DT_D}})]}\}} \\
+ (1-\zeta_{\mathrm{DT}_{i}}) \cdot 1 / \{1+\exp{[\beta^{*}_{\mathrm{DT}_{i}}(V_{ \mathrm{PL} }-V_{0_{i},\mathrm{{DT}_\mathrm{T}}})]}\}
\label{eq:combined}
\end{multline}
where $\zeta_{\mathrm{DT}}$ is weight of the non-equilibrium decay-cascade component in the combined model, having statistical bounds of confidence interval between 0 and 1, $\alpha^{*}_{\mathrm{DT}}$ is the temperature-independent gate-referred tunnel rate constant obtained from decay-cascade fit, $1-\zeta_{\mathrm{DT}}$ is weight of the equilibrium thermal component in the combined model, $\beta^{*}_{\mathrm{DT}}$ is the gate-referred modified thermodynamic beta of the thermal component in the combined model, and $V_{0,\mathrm{DT}_\mathrm{D}}$ and $V_{0,\mathrm{DT}_\mathrm{T}}$ are the threshold voltage of decay-cascade and thermal component in the combined model, respectively.

To investigate the dependency of temperature on the charge pumping mechanism, we fit the $\langle n \rangle$ associated with one and two electron pumping plateau, measured as a function of $V_{\mathrm{PL}}$ at varied $T_{\mathrm{base}}$, while keeping the top gate DC voltages constant at $V_{\mathrm{BL}}$ = 1.20~V, $V_{\mathrm{BR}}$ = 2.28~V, $V_{\mathrm{SL}}$ = 2.2~V, $V_{\mathrm{DL}}$ = 2.2~V, $V_{\mathrm{C1}}$ = 0~V and $V_{\mathrm{C2}}$ = 0~V. Fitting of the measured data to the aforementioned decay-cascade (in Eq. \ref{eq:decaycascade}), thermal (in Eq. \ref{eq:thermal}) and combined (in Eq. \ref{eq:combined}) models, are shown in Fig.~\ref{fig:schematic}(d). The magnified-axes in the insets of Fig.~\ref{fig:schematic}(d), visibly depict the decay-cascade model fits better at lower $T_{\mathrm{base}}$. However, with an increment of $T_{\mathrm{base}}$ the measured plateaus starts agreeing more with the thermal model. To quantify the electron transfer mechanism as a function of $T_{\mathrm{base}}$ we implement two approaches. First, individually determining the average residual sum of squares of the decay-cascade $(\mu_{RSS,\mathrm{D}})$ and thermal $(\mu_{RSS,\mathrm{T}})$ fit to the experimental data. A statistical value of $\Delta \mu_{RSS}$ $ (=\mu_{RSS,\mathrm{D}}-\mu_{RSS,\mathrm{T}})$ less than zero implies a better fit to the decay-cascade model, whereas a positive $\Delta \mu_{RSS}$ indicates the paramountcy of the thermal model, in Fig.~\ref{fig:1e&2e}(a). Second, we benchmark the weight component of $\zeta_{\mathrm{DT}}$ at 0.5, signifying the transition from the regions where decay-cascade or thermal model fit better, in Fig.~\ref{fig:1e&2e}(b). The operating regime of charge pump is purely decay-cascade if $\zeta_{\mathrm{DT}}$ is about unity, in contrast estimation of $\zeta_{\mathrm{DT}}$ close to zero, corresponds to thermal mechanism of electron capture in the QD. From the results in Fig.~\ref{fig:1e&2e}(a, b), we find our charge pump operates in decay-cascade regime at $T_{\mathrm{base}}$ less than 5~K while transferring single-electron per voltage cycle. However, pumping two electrons shifts this decay-cascade to thermal transition below a temperature of 4~K.

The phenomenological fit parameter $\alpha^{*}_{\mathrm{D}}$ relate to tunneling rate of excess electrons $(\Gamma^{\mathrm{escape}}_{\mathrm{D}_{i}})$, escaping back to the source reservoir leaving $\langle n \rangle$ number of electrons in the QD. One can theoretically forecast the lower bound of uncertainties encountered during the process of pumping from the difference $(\alpha^{*}_{\mathrm{D}_{2}}\cdot V_{0_{2},\mathrm{D}}) - (\alpha^{*}_{\mathrm{D}_{1}}\cdot V_{0_{1},\mathrm{D}})$, equivalent to the difference of back tunneling rate of excess electrons $\ln{\Gamma^{\mathrm{escape}}_{\mathrm{D}_{2}}} - \ln{\Gamma^{\mathrm{escape}}_{\mathrm{D}_{1}}}$~\cite{kashcheyevs2010universal,yamahata2014accuracy}. Although $\alpha^{*}_{\mathrm{D}}$ is independent of temperature, the regression analysis results deduced a decrement in $\alpha^{*}_{\mathrm{D}_{1}}$ and $\alpha^{*}_{\mathrm{D}_{2}}$, when the charge pump is operated in the thermal regime, corresponding to a $T_{\mathrm{base}}$ higher than the transition temperature. In contrast, we observe saturation of $\alpha^{*}_{\mathrm{D}}$ for both one and two electron transfer, when $T_{\mathrm{base}}$ is lower than the transition temperature, in Fig.~\ref{fig:1e&2e}(c).

Although the charge pump is cooled to $T_{\mathrm{base}}$ temperature, the local electron temperature in the source and drain reservoir leads $(T_{\mathrm{pump}})$ may be higher due to AC periodic drive induced heating. In order to assess, the $T_{\mathrm{pump}}$ whilst pumping one and two electrons per voltage cycle, we infer the gate-referred modified thermodynamic beta extracted from the thermal component of the combined model to evaluate the value of $T_{\mathrm{pump}_{i}} = (e \cdot \alpha_{\mathrm{PL-QD}})/(k_{B} \cdot \beta^{*}_{\mathrm{DT}_{i}}) $. To pursue a fair comparison of local electron temperature at the source reservoir as a function of varying $T_{\mathrm{base}}$, we evaluate the heat induced in the pump $\Delta T_{\mathrm{pump}} = T_{\mathrm{pump}}-T_{\mathrm{base}}$, in Fig.~\ref{fig:1e&2e}(d). We assess, the heat induced due to AC periodic drive in our system is higher during transferring single-electron per cycle, when compared to that of two electrons transfer. However, the number of electrons existing in the QD before initialization of charge pumping proccess might have an influence in the value of $\Delta T_{\mathrm{pump}}$. Therefore, the gate-referred modified thermodynamic beta used to deduce the $T_{\mathrm{pump}}$ for two electrons plateau deserves further theoretical and experimental investigation.

\begin{figure}[hbt!]
\includegraphics[width=0.5\textwidth]{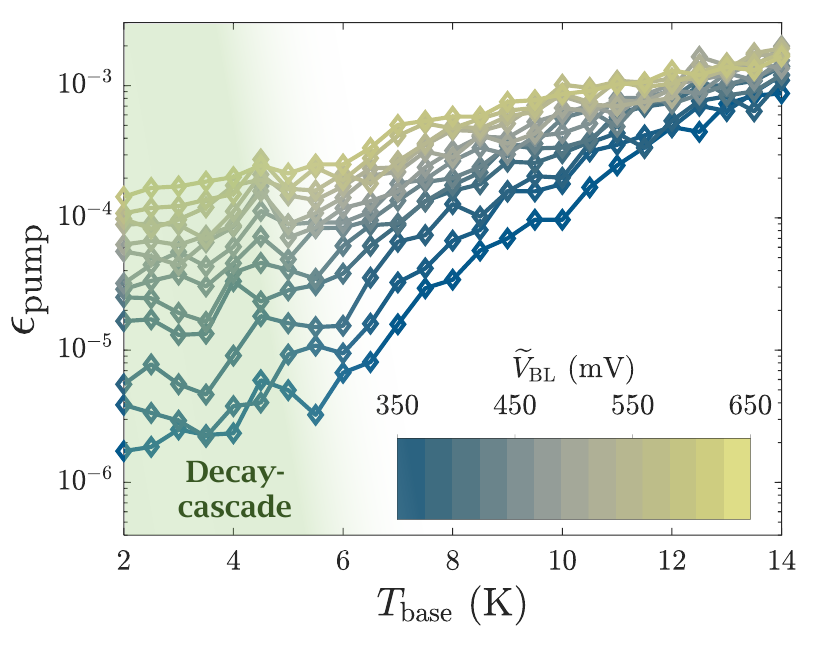}
\caption{\label{fig:calculated_error} Lower bound of single-electron pumping error $(\epsilon_{\mathrm{pump}})$ as a function of varying  drive amplitude $(\widetilde{V}_{\mathrm{BL}})$ and base temperature of sample-space ($T_{\mathrm{base}}$).}
\vspace{-1.5em}
\end{figure}

Next, we turn our attention towards the impact of $\widetilde{V}_{\mathrm{BL}}$ on the single-electron pumping fit parameters ($\Delta \mu_{RSS}$, $\zeta_{\mathrm{DT}}$, $\alpha^{*}_{\mathrm{D}}$ and $\Delta T_{\mathrm{pump}}$) as a function of $T_{\mathrm{base}}$ and $\widetilde{V}_{\mathrm{BL}}$. The top-gate DC voltages are kept constant at $V_{\mathrm{BL}}$ = 1.32~V, $V_{\mathrm{BR}}$ = 2.38~V, $V_{\mathrm{SL}}$ = 2.2~V, $V_{\mathrm{DL}}$ = 2.2~V, $V_{\mathrm{C1}}$ = -0.03~V and $V_{\mathrm{C2}}$ = 0~V while repeating the measurement for different values of $\widetilde{V}_{\mathrm{BL}}$ and $T_{\mathrm{base}}$. The $\Delta \mu_{RSS}$ and $\zeta_{\mathrm{DT}}$ in Fig.~\ref{fig:1e&2e}(e) and Fig.~\ref{fig:1e&2e}(f), respectively, deduce a leftward shift of the transition temperature with rising $\widetilde{V}_{\mathrm{BL}}$. In addition, it is worth noting the value of  $\alpha^{*}_{\mathrm{D}}$ deteriorates with increment in $\widetilde{V}_{\mathrm{BL}}$ in Fig.~\ref{fig:1e&2e}(g). This informs us the importance of tuning the $\widetilde{V}_{\mathrm{BL}}$ amplitude. Further, the evaluated value of $\Delta T_{\mathrm{pump}}$ support the statement as at a particular $T_{\mathrm{base}}$ the heat induced is directly proportional to $\widetilde{V}_{\mathrm{BL}}$, in Fig.~\ref{fig:1e&2e}(h).

\vspace{0.7em}
In order to pursue a qualitative investigation of the single-electron pumping uncertainties, we follow a theoretical approach to assess the lower bound of error occurring during the single-electron pumping processes~\cite{kashcheyevs2010universal}. To determine the charge pumping error $\epsilon_{\mathrm{pump}}$ as a function of $\widetilde{V}_{\mathrm{BL}}$ and $T_{\mathrm{base}}$, we use the single-electron transfer experimental data, which is measured as a function of $V_{\mathrm{PL}}$ by varying $\widetilde{V}_{\mathrm{BL}}$ and $T_{\mathrm{base}}$, a single sweep of whose (at $\widetilde{V}_{\mathrm{BL}} = $ 400) is shown in Fig.~\ref{fig:schematic} (c). The $\epsilon _{\mathrm{pump}}$ is given as $1 - \langle n_{\mathrm{D}} \rangle$, at the point of inflection $(V_{\mathrm{PL}}^*)$ on $\langle n \rangle =1$ plateau. While calculating the $\epsilon_{\mathrm{pump}}$, we assumed $\alpha^{*}_{\mathrm{D}_{1}} = \alpha^{*}_{\mathrm{D}_{2}}$ and $\Delta V_{0,\mathrm{D}} (= V_{0_{2},\mathrm{D}}-V_{0_{1},\mathrm{D}})$ is independent of $\widetilde{V}_{\mathrm{BL}}$. The lower bound of the evaluated single-electron pumping error as a function of $T_{\mathrm{base}}$ and $\widetilde{V}_{\mathrm{BL}}$ is illustrated in Fig.\ref{fig:calculated_error}. The theoretical  error rate of our Si single-electron pump when operated at a $T_{\mathrm{base}}$ of 2 K and $\widetilde{V}_{\mathrm{BL}} =$ 350 mV is 1.72 ~ppm. However, the theoretical uncertainty figure might be overestimated compared to our charge pump capability due to the normal accuracy measurements using  the voltmeter~\cite{giblin2012towards}. At a constant $T_{\mathrm{base}}$, the reduction in $\widetilde{V}_{\mathrm{BL}}$ lead to sharper edges of the plateau, and theoretically lower error bound. Besides, the captivating results delineate, single-electron pumping error is almost independent of $T_{\mathrm{base}}$ when operated in a regime where an electron is captured in the pump QD by following a sequence of tunneling event back to the source reservoir.

Overall, we showed that our Si single-electron pump is operable at liquid helium temperature with tunnel limited errors dominating the pump fidelity, indicating high precision metrological current measurements could be consistently done in cheap $^{4}$He systems. This pave the path for transportable and scalable primary SI current standard by deploying SiMOS technology, which is well-established across the semiconductor foundries. To future characterise the charge pumping errors we will use an on-chip charge sensor.

A.D. performed all measurements and all calculations under T.T.'s supervision. S.Y. and F.H. fabricated the device under A.S.D's supervision. A.D., S.Y., M.K.F., A.S., and T.T. participated in data interpretation. A.D., S.Y., and T.T. designed the project and experimental setup. A.D. wrote the manuscript with contribution from all authors.

We thank Md. Mamunur Rahman and Alexandra Dickie for assistance in the cryogenic setup. We acknowledge support from the Australian Research Council (DP200103515), the U.S. Army Research Office (W911NF-17-1-0198), and NSW node of the Australian National Fabrication Facility. A.D., and M.K.F. acknowledge scholarship support from the Sydney Quantum Academy, Australia.

\bibliography{ChargePumpingP1_References.bib}
\end{document}